# A chemically peculiar Be-shell star in a sub-solar metallicity solution for the post-mass-transfer eclipsing binary V658 Car

Norbert Hauck

Bundesdeutsche Arbeitsgemeinschaft für Veränderliche Sterne e.V. (BAV), Munsterdamm 90, 12169 Berlin, Germany; hnhauck@yahoo.com

**Abstract:** V658 Car (HD 92406) is a newborn resp. rejuvenated shell star binary system at the age of only 1 Myr after the end of mass transfer. In this renewed study the peculiarities of the Be star are at first by-passed by the determination of the properties of the contracting hot subdwarf precursor, but finally resolved by combining photometric data and radial velocity results with existing stellar models. For the effective temperatures, radii and masses we get about 12900 K, 1.92/2.20 Rsun and 4.49 Msun for the Be star, and about 18400 K, 1.87 Rsun and 0.56 Msun for its companion star. The Be star has a rotational velocity of 336 km/s and is surrounded and dimmed in our view by a large and luminous equatorial decretion disk having a radius of ~ 42 Rsun. According to stellar models these results fit to a surprisingly low metallicity Z of 0.003 and a $T_{eff}$ ~ 5400 K higher than observational expectations for the Be star, which hence should belong to the chemically peculiar stars, in spite of its rapid rotation.

V658 Car (HD 92406) is the first found exemple of an eclipsing shell star binary system and has been studied already twice by the author [1,2]. In the meantime we have received new data from the TESS mission as well as results from other sources. Moreover, we are now using a revised approach to this complicated system comprising three different sources of light, i.e. a significantly dimmed and peculiar shell star, a luminous decretion disk and a post-mass-transfer remnant of the former donor star. Instead of trying to fit the primary Be star to known *peculiar* spectral types (e.g. A0p or (B5V)p shell), we this time avoid any confusion by the variable composite spectra of the binary system and investigate first of all the light of the secondary star.

A detailed light curve of V658 Car from TESS data is shown in Fig. 1. The stellar partial eclipses in phases 0 and 0.50 are embedded in larger funnel-shaped dimming effects of the decretion disk. The disk and primary stellar eclipses around phase 0 both are attributed *exclusively* to the loss of light of the **secondary star**, when it is dimmed by the large disk and partially eclipsed by the primary star.

Our photometric *UBV*-band data obtained by a remotely controlled CDK 20-inch telescope in Siding Spring, Australia, have been used for the light curve modeling with the *Binary maker 3* (BM3) software tool. The geometrical stellar parameters values have been derived mainly from the best primary minimum fit in the *U*-band, which achieved a $\sigma_{FIT}$ of 3.9 mmag, and could remain nearly unchanged against those published and depicted in Fig. 5 of our last V658 Car paper [2]. The primary stellar eclipse has been transformed with the BM3 tool from a 52.8 % partial into a total one, and the depths obtained have been added to those we have observed for the disk eclipse between maximum light and first/last contact of the primary stellar eclipse in order to get the total light of the secondary star in the *UBV*-bands (see Table 2). The apparent magnitudes of V658 Car at maximum light have been taken from the literature, i.e. 9.13 *V*mag from Fernie [3], and 9.10 *B*mag (via *B*–*V* = – 0.03) as well as 8.57 *U*mag from Deutschman et al. [4]. These *UBV*-Magnitudes appear to be correct, and avoid an inacceptable final result for the interstellar extinction $A_V$ of < 0. From Table 3 of Wang & Chen [5] the coefficients $A_U$ = 1.584 $A_V$ and $A_B$ = 1.317 $A_V$ for normal interstellar extinction have been used. The obtained color indexes of $(U – B)_0$ and $(B – V)_0$ in combination with table 5 of Pecaut & Mamajek [6] finally give an effective temperature $T_{eff}$ of 18360 ± 500 K for the secondary star as well as an interstellar extinction $A_V$ of 0.01 mag.

This secondary star then has been fitted to the evolutionary tracks of post-mass-transfer remnants of 0.378 $M_\odot$ and 0.523 $M_\odot$ depicted in the Hertzsprung-Russell (HR) diagram of Iben & Tutukov [7] in their Fig. 24. The mass-luminosity function derived therefrom has been corrected for our sub-solar metallicity Z ≈ 0.003 (see here below) using the metallicity-dependent HR-diagrams of Istrate et al. [8] (see their Fig. 9 for a post-mass-transfer remnant of 0.28 $M_\odot$). The incomplete radial velocity (RV) data of Gieseking [9] used and shown in Fig. 6 of our V658 Car paper [2] have been replaced by the much better determined secondary star's RV semi-amplitude $K_2$ = 102.1 ± 0.2 km s$^{-1}$, which has been selected from a recent publication of de Amorim et al. [10].

Herewith a secondary mass of 0.558 $M_\odot$ fits to a bolometric luminosity $L_2$ of 356 $L_\odot$ and a radius $R_2$ of 1.87 $R_\odot$. The $R_2/a$ ratio known from our geometrical fit gives a separation *a* ≈ 73.0 $R_\odot$ and (via Kepler III) the total mass. The resulting primary mass $M_1$ = 4.49 $M_\odot$ and $M_2$ then simultaneously comply with the mass function calculated with this $K_2$. Detailed results are given in Tables 1 and 2.

The properties of the secondary star are in line with a contracting hot subdwarf precursor. According to Fig. 24 of Iben & Tutukov [7] we can expect a binary age of not more than 1 Myr after the end of Roche-lobe overflow, and an initial donor-star mass of ~ 4 $M_\odot$. Consequently, for the initial mass of the accretor star only ~ 1 $M_\odot$ is left over. For such an extreme initial mass ratio and a post-mass-transfer age of only 1 Myr the newborn Be star should be located still very close to the zero-age main sequence (ZAMS) in the Hertzsprung-Russell diagrams of stellar models.

The size of the **decretion disk** has been derived from the duration of its secondary star eclipse. With a mean radius of ~42 $R_\odot$ it fills out ~98 % ot the Roche limit, however, its eccentric position w.r.t. the primary star might have shifted it locally towards this limit. The disk is expected to be luminous in its central, pseudo-photospheric part, and should have at maximum light a flux fraction of 40.7 % in the *V*-band in order to equalize the calculated photometric distances (~1021 pc) for both stars.

In the light curves of all available TESS data of V658 Car a permanent oscillation having a peak-to-peak amplitude of ~3 to 13 mmag is visible (see Fig. 1-4). The Lomb-

Scargle periodograms show only two prominent oscillation periods, mainly at 0.332(6) days (Fig. 2,3), and sometimes at half thereof at 0.163(2) days (Fig. 4). The dominant, longer period corresponding to a rotation rate factor $F_1$ of 97 in V658 Car's 32-days orbit has successfully reproduced the best solution, e.g. a $\sigma_{FIT}$ of 3.9 mmag for our photometric *U*-band data of the primary minimum, after only *minor* adjustments to the geometrical input ($F_1$, *i*, et al.) of our last paper [2]. Therefore, this is considered to be the *rotational* period of a *chemically peculiar* Be star, and not a pulsational period. The shorter period is regarded as being caused by occasionally appearing additional chemical patches on the opposite side of the star, as a consequence of a dipolar magnetic field in the Be star, with the axis of the magnetic field being inclined to the axis of rotation, as known from the *oblique rotator model*.

For the **primary star** we then get a flattening ratio of 1.145. The equatorial/orbital rotational velocity ratio W of 336/624 ≈ 0.54 is somewhat below the lower limit of ~0.60 shown in Fig. 9 of Rivinius et al. [11], however, at a rotational/critical velocity ratio $v_{Rot}/v_{crit}$ of 0.62 well inside the revised ranges from 0.3 to 0.95 for Be stars according to Zorec et al. [12]. The mean $T_{eff}$ of the primary shell star in our equatorial view has been determined by comparison with its hotter companion in the light curve fit, and amounts from 11360 K in the secondary minimum, when it is assumed to be dimmed by a tidally locked densification in the decretion disk, to ~11625 K at maximum light, when it is dimmed by the undensified disk. The dimming effect of the entire decretion disk has been measured at the disk eclipse of the secondary star, and then calculated for half of the disk size. This gives a mean $T_{eff}$ of ~12470 K for the Be star without disk in our view, and ~12890 K over its total surface area according to the BM3 output for a comparable sphere. Finally, with help of Fig. 2, 4 and 5 of Ekström et al. [13], we get a theoretical $T_{eff}$ of ~13550 K for the non-rotating spherical star.

Our primary star's theoretical radius of 1.944 $R_\odot$ for a non-rotating star and its mass of 4.49 $M_\odot$ fit, as required, close to the ZAMS between non-rotating stellar models of Georgy et al. [14] at a surprisingly low sub-solar metallicity Z of 0.003 at a $T_{eff}$ of ~19000 K, i.e. ~5450 K higher than our observationally based calculation, which confirms the *chemically peculiar* Be star finding here above.

According to the available literature (e.g. *CDS* catalogue J/MNRAS/468/2745 from Netopil et al. [15]) V658 Car's Be star showing rapid equatorial rotation of ≈ 336 km s$^{-1}$ at a rotational period of only ≈ 0.332 days is among the chemically peculiar (CP) stars one of the fastest rotators found to date. In comparison with known CP stars, this Be star is compensating its increased centrifugal by a higher gravitational acceleration resp. mass/radius ratio. Thereby, the chemical patches at the stellar surface are *not* being erased by rotational mixing.

**Acknowledgements :** This research has made use of the Simbad and VizieR databases operated at the *C*entre de *D*onnées astronomique, *S*trasbourg, France. This paper also includes data collected with the TESS mission, obtained from the MAST data archive at the Space Telescope Science Institute (STScI).

**Table 1 : Parameters of the binary system V658 Car**

| | | |
|---|---|---|
| Epoch [HJD] | 2452786.438(2) | mid primary minimum (phase 0) |
| Orbital period [days] | 32.18537(2) | including TESS data |
| Apparent $V$ magnitude | 9.13 | from Fernie [3] |
| Apparent $B$ magnitude | 9.10 | from [3] + [4] |
| Apparent $U$ magnitude | 8.57 | from Deutschman et al. [4] |
| Eclipse duration [hours] | 12.36 | stellar eclipses |
| Eclipse duration [days] | 6.495(10) | disk eclipse |
| Orbital inclination $i$ [deg] | 88.75 ± 0.10 | and circular orbit adopted |
| Semi-major axis $a$ [$R_\odot$] | 72.97 ± 0.15 | for $R_\odot$ = 696342 km |
| Distance [pc] | 1021 ± 22 | photometrical |
| Extinction $A_V$ [mag] | 0.01 | interstellar |

**Table 2 : Parameters of the components of V658 Car**

| Parameter | primary star | secondary star | disk |
|---|---|---|---|
| Radius (mean) [$R_\odot$] | 2.11 ± 0.03 | 1.87 ± 0.03 | 42 |
| Radius (pole/equator) [$R_\odot$] | 1.923 / 2.202 | | |
| Rotational velocity [km s$^{-1}$] | 336 ± 6 | | |
| Temperature mean $T_{eff}$ [K] | 12890 ± 350 | 18360 ± 500 | |
| $V$-flux fraction at max. light | 0.207 | 0.386 | 0.407 |
| $B$-flux fraction at max. light | | 0.449 | |
| $U$-flux fraction at max. light | | 0.536 | |
| Apparent $V$ magnitude | 10.841 | 10.163 | |
| Apparent $B$ magnitude | | 9.970 | |
| Apparent $U$ magnitude | | 9.246 | |
| Luminosity (bolometric) [$L_\odot$] | 109 ± 13 | 356 ± 41 | |
| Mass [$M_\odot$] | 4.49 ± 0.03 | 0.558 ± 0.005 | |

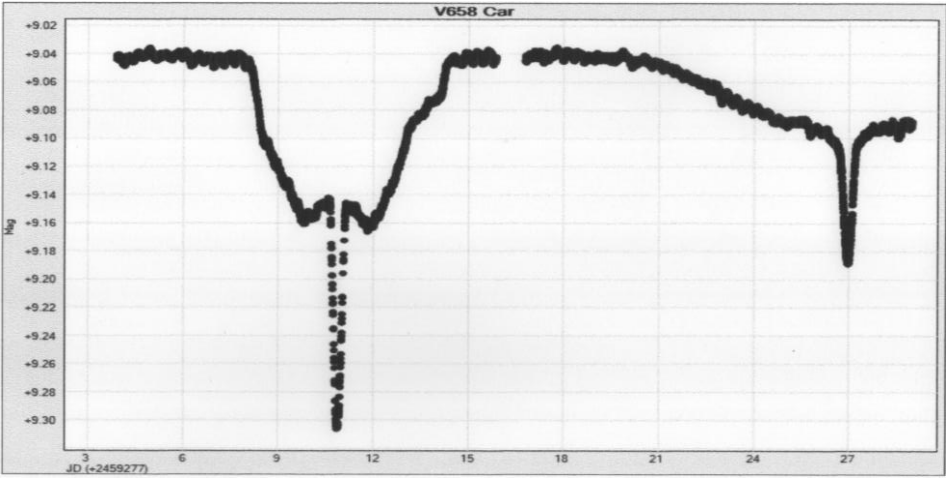

**Fig. 1:** TESS light curve of V658 Car (march 2021) from phase 0.775 to 0.565 of the 32-days orbit. Note the dimming effects of the disk, stellar eclipses and the oszillations.

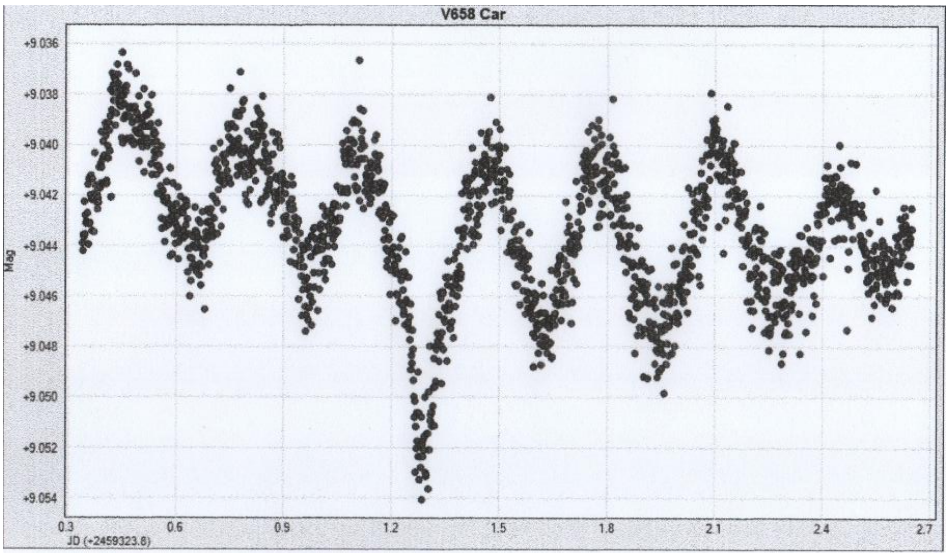

**Fig. 2:** TESS light curve of V658 Car (april 2021) from phase 0.126 to 0.198 showing a rotational period of 0.330 days for an obviously large (hemispherical) chemical patch around one of the magnetic poles.

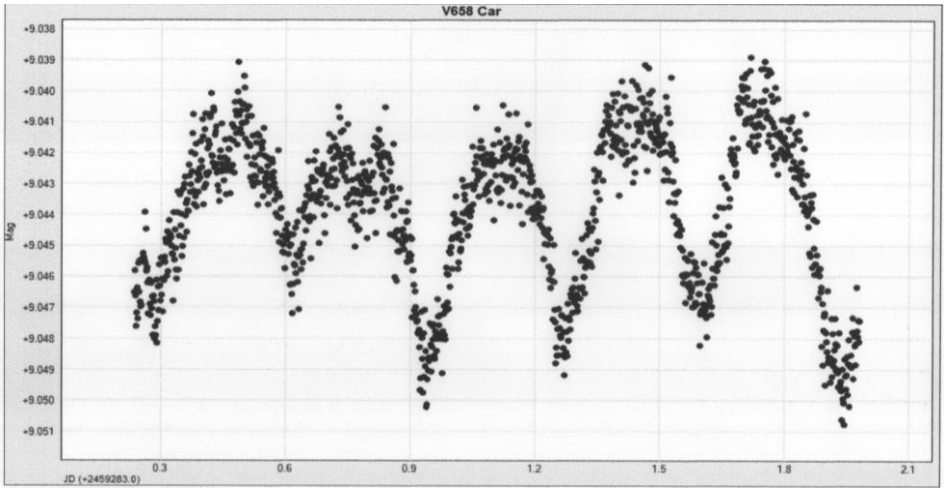

**Fig. 3:** TESS light curve of V658 Car (march 2021) from phase 0.856 to 0.910 showing a rotational period of 0.328 days for an apparently smaller chemical patch around one of the magnetic poles. The shape of the maxima is flattened, when this patch is not visible for us.

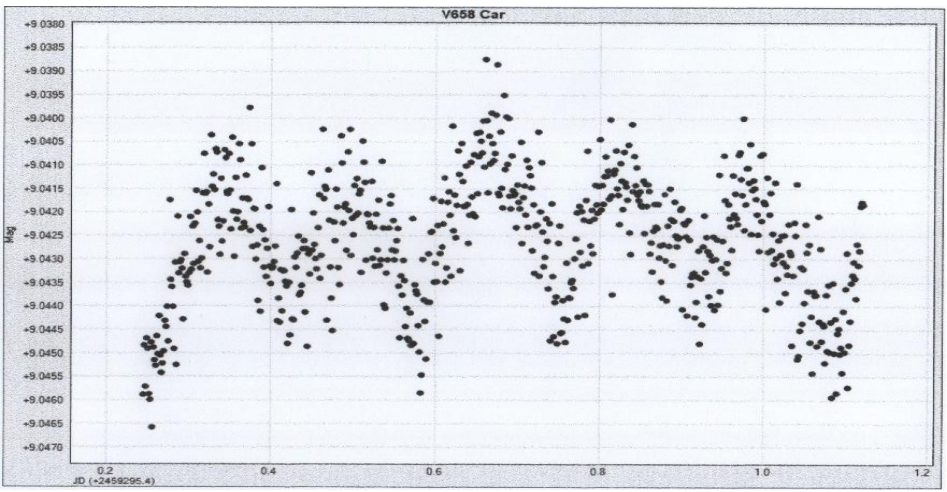

**Fig. 4:** TESS light curve of V658 Car (march 2021) from phase 0.241 to 0.268 showing a rotational period of 0.163/0.164 days for smaller chemical patches around both of the magnetic poles on opposite sides of the Be star.